# Lattice-mismatched semiconductor heterostructures

Dong Liu[1†], Sang June Cho[1†], Jung-Hun Seo[1†], Kwangeun Kim[1], Munho Kim[1], Jian Shi[2], Xin Yin[2], Wonsik Choi[3], Chen Zhang[3], Jisoo Kim[1], Mohadeseh A. Baboli[4], Jeongpil Park[1], Jihye Bong[1], In-Kyu Lee[1], Jiarui Gong[1], Solomon Mikael[1], Jae Ha Ryu[1], Parsian K. Mohseni[4], Xiuling Li[3], Shaoqin Gong[5], Xudong Wang[2], and Zhenqiang Ma[1*]

[1]Department of Electrical and Computer Engineering, University of Wisconsin –Madison, Madison, WI 53706, USA

[2]Department of Materials Science and Engineering, University of Wisconsin –Madison, Madison, WI, USA

[3]Department of Electrical and Computer Engineering, University of Illinois – Urbana Champaign, Champaign, IL 61820, USA

[4]Microsystems Engineering and NanoPower Research Laboratories, Rochester Institute of Technology, Rochester, NY 14623, USA

[5]Department of Biomedical Engineering, University of Wisconsin –Madison, Madison, WI, USA

[†]These authors contributed equally.

[*]Correspondence should be addressed to: mazq@engr.wisc.edu

**Semiconductor heterostructure is a critical building block for modern semiconductor devices. However, forming semiconductor heterostructures of lattice-mismatch has been a great challenge for several decades. Epitaxial growth is infeasible to form abrupt heterostructures with large lattice-mismatch while mechanical-thermal bonding results in a high density of interface defects and therefore severely limits device applications. Here we show an ultra-thin oxide-interfaced approach (named as "grafting") for the successful formation of lattice-mismatched semiconductor heterostructures. Following the depiction of a theory on the role of interface oxide in forming the heterostructures, we describe experimental demonstrations of Ge/Si (diamond lattices), Si/GaAs (zinc blende lattice), GaAs/GaN (hexagon lattice), and Si/GaN heterostructures. Extraordinary electrical**

**performances in terms of ideality factor, current on/off ratio, and reverse breakdown voltage are measured from p-n diodes fabricated from the four types of heterostructures, significantly outperforming diodes derived from other methods. Our demonstrations indicate the versatility of the ultra-thin-oxide-interface approach in forming lattice-mismatched heterostructures, open up a much larger possibility for material combinations for heterostructures, and pave the way toward broader applications in electronic and optoelectronic realms.**

One of the most important inventions in the last century that has deeply impacted human society [1] is the semiconductor heterostructure [2-6]. To form a heterostructure, lattice match (or a very small mismatch) between two adjacent semiconductor materials that allows epitaxial growth of one on top of the other is an essential requirement [7]. Restricted by the stringent requirement, only a few systems of heterostructures can be formed via epi-growth techniques. Despite the great success of lattice-matched heterostructures, forming lattice-mismatched heterostructures has been sought after for over six decades, yet without a solution found.

Among the many attempts to form lattice-mismatched heterostructures, the most notable approaches are heteroepitaxy [8-12] and mechanical-thermal direct bonding (or fusion) of two materials [13-17]. When two single crystalline materials are directly grown or fused together, a large number of defects at the interface are inevitably generated due to lattice mismatch. The high interface traps induced by these defects act as recombination and generation centers which severely degrade charge carrier transport at the interface. The poor interface quality, which is much worse than that of lattice-matched heterostructures grown epitaxially, renders these two approaches impractical for most device applications, despite demonstrations of some useful devices using the latter approach [13, 17] where only the interface electrical conductivity is of concern.

Considering that two crystalline semiconductors cannot be directly combined to form an interface of low density of states, the two materials ought to be physically separated. In order to allow charge carriers to transport across the interface with minimal loss, which necessitates a sufficiently low density of recombination/generation centers at the interface, a third material (denoted as the "interface layer") that satisfies the following requirements needs to be inserted in between. First, it must have the ability to effectively passivate the surfaces of both semiconductor materials. Second, it must allow both electrons and holes to cross with minimal loss. Third, it can prevent interdiffusion of the two connected neighboring materials during thermal processing and junction operation. Fourth, it can provide sufficient adhesion force for the two materials. And lastly, it can act as a thermal expansion buffer when the materials are subject to elevated temperatures, e.g., for processing.

To satisfy the above requirements, metallic materials are out of the question because they are gapless materials and holes cannot pass a metal layer. The only candidate left for consideration as the potential interface material is insulators, such as oxides, nitrides, *etc*. It is known that many types of oxides can effectively passivate various semiconductor surfaces [18-25]. Since oxides and nitrides have a bandgap, electrons and holes can indeed exist in these materials. They are considered insulators because of their ultra-wide bandgap and few intrinsic charge carriers exist inside them at room temperature. However, when oxides and nitrides are made ultra-thin, charge carriers can pass through via quantum tunneling, which has been known for decades [26-28]. While it is well-known that oxides and nitrides can function as passivation materials and their ultrathin form can also function as a tunneling layer, whether their ultrathin form can function as both a passivation and a tunneling layer in between two lattice-mismatched single crystalline semiconductors such that equivalent-to-epitaxial-junction heterojunctions can be realized is unknown and no relevant studies have been found in existing literature.

In this paper, we report using ultrathin oxide ($Al_2O_3$) as the interface material to overcome the lattice-mismatch challenges in forming semiconductor heterostructures. We first present an interface theory model to formulate an analytical relation between the ideality factor (*n*) of a heterostructure p-n junction diode and interface density of states of the heterostructure. Then, we use the Ge/Si (diamond lattice/diamond lattice) heterostructure as an example to detail the fabrication process and interface information, and to analyze the characteristics of Ge/Si p+/n-diodes. We further describe Si/GaAs (zinc blende lattice), GaAs/GaN (hexagon lattice), and Si/GaN heterostructures, with a lattice mismatch of 4.2%, 4.9%, 77.1%, and 70.2%, respectively, that were formed employing the same method to demonstrate the versatility of the approach.

**Theory Model**

Figure 1 illustrates how the surface states and interface layer jointly influence the energy band alignment and charge carrier transport within a generic heterostructure p-n junction. With an interface layer inserted in between two semiconductors, two interfaces between the interface layer material and the two semiconductors are created at the junction. The surface charge densities at the two interfaces depend on how well each of the two surfaces of semiconductors is passivated. The charges on the two sides of the interface layer induces a voltage drop across it. For the p-n junction shown in Figure 1a, its net charge distribution is sketched in Figure 1b. To facilitate the analysis, a few reasonable assumptions are made (see SI) along with the depletion approximation. Accordingly, the electric field intensity $E(x)$ and the corresponding potential $\phi(x)$ are schematically illustrated in Figure 1c and Figure 1d, respectively. Due to the existence of an electric field within the interface layer, the potential change for aligning the two Fermi levels, $E_{F1}$ and $E_{F2}$, (Figure 1e) in the p-n junction is

$$\Delta E_F = \psi_p + V_i + \psi_n \qquad (1)$$

where $\psi_p$, $\psi_n$, and $V_i$ are the built-in voltages on the p side and n side, and voltage drop on the interface layer, respectively.

Without violating any general principles, in the following analysis, we use a one-sided p+/n-Ge/Si junction ($N_a \gg N_d$) (Fig. 1e) as our example to demonstrate the influence of the interface layer on band alignment. The energy band discontinuities at conduction and valence band, defined by the electron affinity difference, are $\Delta E_{c0}$ and $\Delta E_{v0}$, respectively. After the two materials join with the interface layer ($Al_2O_3$) in between, the Fermi level is flat to maintain an equilibrium state as shown in Figure 1f. Due to the voltage drop on the interface layer (Fig. 1d), the band discontinuities at conduction and valence bands become:

$$\Delta E_c = \Delta E_{c0} + qV_i \quad ; \quad \Delta E_v = \Delta E_{v0} - qV_i \qquad (2)$$

If the interface layer effectively suppresses the surface states by passivation, the direct tunneling process [29, 30], rather than traps-assisted tunneling [31], ought to be the primary charge carrier transport mechanism through the interface layer. Considering the relatively large abrupt energy band discontinuity in our heterostructure, the thermionic emission model is the suitable mechanism for current flow across the heterostructure interface [32, 33].

When a forward bias $V$ is applied to the p-n junction, the built-in potentials $\psi_p$ and $\psi_n$ are reduced by $\Delta\psi_n$ and $\Delta\psi_p$, due to the changes of $Q_{scn}$ (i.e., $\Delta Q_{scn}$) and $Q_{scp}$ (i.e., $\Delta Q_{scp}$) (Fig. 1b), respectively, to facilitate the charge carrier transport as shown in Figure 1g. $\Delta\psi_n$ and $\Delta\psi_p$ also subsequently alter the occupation probabilities of the surface states for the semiconductors on both sides, inducing changes of $Q_{ssn}$ (*i.e.*, $\Delta Q_{ssn}$) and $Q_{ssp}$ (*i.e.*, $\Delta Q_{ssp}$) (Fig. 1b). Together, a change in $V_i$, $\Delta V_i$, is induced, leading to $V = \Delta\psi_n + \Delta\psi_p + \Delta V_i$. Since it is the $\Delta\psi_n$ and $\Delta\psi_p$ that control current transport rather than $V$ based on the thermionic emission model [34], $n$ is derived as

$$n = \frac{V}{\Delta\psi_n + \Delta\psi_p} = 1 + \frac{\Delta V_i}{\Delta\psi_n + \Delta\psi_p} \quad (3)$$

For a p+/n- junction, $\Delta\psi_p$ is negligible compared to $\Delta\psi_n$. Combining $\Delta V_i = W_i \frac{\Delta Q_{scn} + \Delta Q_{ssn}}{\varepsilon_i}$ with the depletion approximation $\Delta Q_{scn}/\Delta\psi_n = \varepsilon_n/W_n$, and $\Delta Q_{ssn} = q\Delta\psi_n D_{ssn}$, by assuming a uniform distribution of surface states across the energy gap, it is found (see SI for the detail) that

$$n = 1 + \frac{W_i}{\varepsilon_i}\left(\frac{\varepsilon_n}{W_n} + qD_{ssn}\right) \quad (4)$$

where $W_i$ and $\varepsilon_i$ are the thickness and permittivity ($\varepsilon_i = \varepsilon_0\varepsilon_{ri}$, $\varepsilon_{ri}$ is relative permittivity of the interface layer material and $\varepsilon_0$ is vacuum permittivity) of the interface layer, respectively. $\varepsilon_n$ is the permittivity ($\varepsilon_0\varepsilon_{rn}$) of the n-type semiconductor. $W_n$ is the depletion width (Fig. 1a) and $D_{ssn}$ (*i.e.*, $D_{it}$, which is commonly found in other publications) is the n-side surface density of states. Figure 1h provides a calculated graphic representation of equation (4).

Since $W_n \gg W_i$, equation (4) can be simplified to $n = 1 + \frac{1.81\times10^{-13}}{\varepsilon_{ri}} W_i(nm) D_{ssn}$ ($\varepsilon_{ri}$ of Al$_2$O$_3$: ~6.74 (Figure S1). If the $D_{ssn}$ value can be reduced to $1.0\times10^{12}$ /eV·cm², $n$ can reach the range of 1.01-1.05 for reasonable tunneling thicknesses of $W_i$ (<1 nm). It has been reported that a $D_{ssn}$ of as low as ~$10^{11}$ /eV·cm² can be reached for Si [35, 36] surfaces, indicating the opportunity of realizing epitaxially grown-like heterostructure p-n junctions with the ultrathin Al$_2$O$_3$ as an interface layer. It is worth noting that equation (4) also indicates the critical importance of suppressing the surface density of states of the light-doped semiconductor side than the more heavily doped side. Furthermore, for a heterostructure diode with $n$ close to unity, the current density levels under reverse bias should be only determined by the properties of the two-sided bulk semiconductors, independent of the heterostructure interface. As a result, such heterostructure diodes are expected to exhibit optimal rectification ratios and behave like epitaxially grown lattice-matched diodes.

**Ge/Si heterostructure**

Although other types of oxides may also be suitable, we adopted $Al_2O_3$ as the interface layer for heterostructure demonstrations because of its excellent passivation capability [20, 35, 36] and outstanding conformal coverage even under very thin thickness using atomic layer deposition (ALD). Figure 2a illustrates the schematic of a p+/n- Ge/Si heterostructure with an ultrathin $Al_2O_3$ as its interface. The detailed heterostructure and diode fabrication process flow can be found in Figure S3. It is noted that the single-crystalline Ge layer in the final heterostructure has no residual strain (Figure S5), despite the large thermal mismatch of Ge and Si. The absence of strain indicates that the interface $Al_2O_3$ served as a thermal buffer layer during the thermal bonding process. Fig. 2b shows a scanning electron microscope image of a finished diode with a Ge layer mesa diameter of 10 μm.

Figure 2c shows a high-resolution transmission electron microscopy (HRTEM) image of the Ge/Si heterostructure interface. The insets are the diffraction patterns of individual bulk material with zone indexing. After forming the heterostructures, the single crystallinity of both Ge and Si were maintained (also see Figure S5). The $Al_2O_3$ in this specifically examined heterostructure was deposited for 5 cycles. The $Al_2O_3$ layer is clearly shown in the HRTEM image as the light-colored region. The widening and slight variation in the width of the light-colored region could be due to the uneven surfaces of Ge and Si, and also possibly due to the slight reflow and/or diffusion of $Al_2O_3$ during thermal anneal. Irrespective of the width variation of the light-colored region, intermixing of Si and Ge did not occur, indicating that the $Al_2O_3$ layer is a diffusion barrier at the anneal temperature. It is worth noting that the $Al_2O_3$ layer shows discernable periodic patterns, which indicate strongly bonded interfaces and the possibility of partial recrystallization. According to the crystal orientations revealed by the HRTEM image, Figure 2d illustrates the atomic-scale schematic of the interface region, in which we attribute the passivation effect of $Al_2O_3$ to the formation of Si-O and Ge-O bonds, even though the $Al_2O_3$ layer is generally in its amorphous form.

The effects of the interface $Al_2O_3$ thickness on heterostructure diode characteristics are examined. For comparison purposes, direct bonding without using an interface layer denoted as 0 cycle was also fabricated as reference devices. Figures 3a-3e show the current density (J) – voltage (V) characteristics of Ge/Si diodes with varying $Al_2O_3$ deposition cycles, 0, 3, 5, 7 and 10 cycles.

For each Al$_2$O$_3$ thickness, multiple devices were measured across the sample area in order to obtain a statistical presentation of the device performance. In comparison with the case of no interface Al$_2$O$_3$ (Fig. 3a), the Ge/Si diodes with 3 (Fig. 3b) and 5 ALD cycles (Fig. 3c) show substantially reduced dark current densities, uniform diode behavior, and exponential forward current characteristics that is well-distinguished from the series resistance-limited region. In contrast, without using Al$_2$O$_3$, the Ge/Si diodes (Fig. 3a) show scattered J-V behavior. Although some diodes in this group also showed low reverse-bias current density, the forward current densities of these diodes are substantially reduced in comparison to the 3- and 5-cycle cases, which in general indicate poor passivation of the interface region. As the Al$_2$O$_3$ deposition is further increased to 7 and 10 cycles, the J-V plots shows deteriorated uniformity, increased ideality factors, and an early onset of resistance-limited region, which are attributed to decreased and/or possibly non-uniform charge carrier tunneling probability across the interface. The best $I_{ON}/I_{OFF}$ ratio at $\pm 1$V for the 3-cycle ALD diode is as high as $1.5 \times 10^8$, which is to date the highest value among state-of-the-art Ge/Si p-n heterojunctions (see Extended Data Table I), including heteroepitaxial growth [37-39] and direct bonding [40-43]. The rapid ramping up of the forward-bias current and the high current density (Figs. 3b and 3c) indicate that there is no significant electric resistance induced by the atomic scale Al$_2$O$_3$ interface layer and the near-unity ideality factor and ultra-low reverse bias current densities of the diodes clearly indicate the sufficient surface passivation of both Si and Ge surfaces.

Figure 3f plots the measured average diode ideality factors of all the diodes, along with their values of deviations (see SI for the deviation calculations) and the corresponding interface states densities, calculated using equation (4), for each group of devices that have the same number of Al$_2$O$_3$ ALD cycles. The average ideality factors for the five cases in Fig. 3a-e are 1.42, 1.02, 1.04, 1.07, and 1.07, respectively. Fig. 3f also reveals that the interface trap densities for all other cases (3-10 cycles) are considerably reduced compared with the case of no interface Al$_2$O$_3$ (direct bonding). The interface states density of the 5-cycle case is estimated to be $7.4 \times 10^{11}$ /eV·cm$^2$, which is comparable with Si metal-oxide-semiconductor (MOS) capacitors that use thicker (>10 nm) Al$_2$O$_3$ as the dielectric [35, 44]. Given the above comparisons, 3~5 cycles Al$_2$O$_3$ deposition can be adopted as the optimal value for realizing Ge/Si heterostructures.

The C-V characteristics of the 5 ALD cycle Ge/Si diodes were measured at room temperature at frequencies of 10 kHz, 100 kHz and 1 MHz and the results are plotted in Figure 3g. The independence of junction capacitance from frequency indicates sufficiently suppressed density of states at the Ge/Si interface. In contrast, the C-V characteristics of the Ge/Si diodes without using $Al_2O_3$ (Figure S7) show a strong dependence on frequency. From Fig. 3g, a flat-band voltage of 0.55 V is extrapolated. In comparison to the theoretical built-in potential (0.46 eV) of a Ge/Si diode based on the electron affinity rule (Figure S8), the deviation of -0.09 eV, which is $qVi$ in equation (2). Taking into account the one-sided p+/n- junction, the majority of band bending potential occurred on the n-type Si side. As a result, the -0.09 eV is manifested in the potential drop across the interface layer with the band up-bending towards the n-Si, as shown in the constructed band diagram (Figure 3h).

Figure 3i depicts the temperature-dependent J-V characteristics of the 5- cycle ALD Ge/Si diode under reverse bias. The Ge/Si diode behaves like epitaxially grown diodes with suppressed current density that slightly increases with the applied reverse bias voltage until an abrupt breakdown that occurs at -26.5 V at room temperature. For the p+/n- junction, the majority of the reverse-bias voltage drop goes to the lightly doped n- Si layer. The breakdown behavior is in distinct contrast to the heterogeneous diodes formed with direct material bonding methods [45], for which high reverse-bias current densities and soft breakdown behaviors are manifested. Furthermore, the breakdown voltages increase from 26.5 V to 27 V and 29 V as the measurement temperature is elevated from 25°C to 50°C and 75°C, respectively. Together, the sharp breakdown behavior, reverse bias voltage sustaining until the bulk Si reaches breakdown, as well as the positive temperature dependence of the breakdown voltage confirm that the breakdown originates from avalanche multiplication within the bulk semiconductors. Due to the effective passivation of the Ge and Si surfaces, it is the generation centers in the Ge and Si bulk regions [46] that are responsible for the reverse-bias current, which is of no difference with lattice-matched diodes formed with epitaxial growth.

**Si/GaAs, GaAs/GaN and Si/GaN heterostructures**

To demonstrate the versatility and more-general applicability of the above-described heterostructure formation method, additional heterostructures, Si/GaAs, GaAs/GaN and Si/GaN, consisting of large lattice-mismatched semiconductor material combinations were formed. For Si/GaAs and Si/GaN, atom inter-diffusion can result in cross doping and contaminations when they make direct contact (without using $Al_2O_3$). The melting temperatures of the three materials are also drastically different besides the large differences in their thermal expansion coefficients (see SI). These factors render excessive challenges in realizing intimate chemical bonding of the two crystalline materials using direct bonding. However, using $Al_2O_3$ of a few monolayers thick as the interface layer seems to solve the bonding challenges as presented in the following. The detailed fabrication processes of the three types of heterostructures and related diodes are illustrated in Figures S9-11. For all three cases, 5 cycles of $Al_2O_3$ were deposited using ALD.

Figures 4a, 4e, and 4i show the HRTEM images of the interface regions of the heterostructures formed between Si/GaAs, GaAs/GaN, and Si/GaN, respectively; and the layer schematics of the three heterostructure diodes are shown as insets of Figures 4b, 4f, and 4j, respectively. The GaN epilayer has a c-plane surface with a hexagonal lattice that is grown on a SiC substrate. The dislocation density of the GaN layer is estimated to be in the order of $10^9/cm^2$ [47]. The insets of Figs 4a, 4e, and 4i are the diffraction patterns of the individual bulk materials with zone indexing. No crystal defects or cracks were observed in the images, indicating the effective function of the interface $Al_2O_3$ as a thermal buffer layer during thermal anneal and cooling. Despite the broadening of the light-colored interface region, inter-diffusion of the two semiconductors in each of the three combinations was not observed. Therefore, the thin $Al_2O_3$ also served as a diffusion barrier layer during the thermal process of heterostructure formation.

Figures 5b, 5f, and 5j show the J-V characteristics of the p+/n- Si/GaAs, p+/n- GaAs/GaN, and p+/n- Si/GaN diodes, respectively. All three types of diodes show practically ideal diode behaviors, *i.e.*, exponential forward current characteristics versus bias voltage and low current under reverse bias. The ideality factors of the Si/GaAs, GaAs/GaN, and Si/GaN diodes are 1.07, 1.13, and 1.14, respectively. The estimated interface trap densities ($D_{ss}$) of the three diodes are $1.1 \times 10^{12}$/eV·cm$^2$, $3.2 \times 10^{12}$/eV·cm$^2$, and $3.5 \times 10^{12}$/eV·cm$^2$, respectively, which are comparable to the values reported from the relevant MOS capacitors and/or surface passivation testing structures [24, 48, 49]. In addition, the three types of diodes all exhibit sharp breakdown characteristics, which are unachievable by other methods, with breakdown voltages of -17.7 V, -30.6V and -28.3V for Si/GaAs, GaAs/GaN, and Si/GaN diodes, respectively (see Extended Data Figure 1). The ideality factor and $I_{ON}/I_{OFF}$ ratio of the three types of diodes are listed in Extended Data Table 1, along with detailed comparisons with the best-performing diodes fabricated using the same material combinations but employing alternative methods. The comparisons illustrate the spectacle of the diodes of this work with the performance metrics of all aspects considerably surpassing the best reported diodes fabricated with wafer bonding [50, 51], wafer fusion [52], surface activated bonding [53], and epitaxy growth [54, 55] techniques. Accordingly, the interface $Al_2O_3$ is proven to have sufficiently suppressed the interface density of states of these heterostructures while providing adequate charge carriers passage, remarkable thermal buffering, and robust adhesion to the bonded materials. The successful demonstrations of these optimal-performance diodes consisting of radically different material combinations have provided strong evidence concerning the wide-ranging applicability of using ultrathin oxide as the interface layer to form lattice-mismatched bulk semiconductor heterostructures. Considering the dramatic differences in the semiconductors used to form the heterostructures and the very different fabrication techniques from epitaxial growth, we name the approach *semiconductor grafting*.

The C-V characteristics of the p+/n- Si/GaAs, p+/n- GaAs/GaN, and p+/n- Si/GaN diodes are characterized and the results are plotted in Figures 5c, Fig. 5g, and Fig. 5k, respectively. The multi-frequency C-V sweeping results are shown in the Extended Data Figure 2, which all reveal nearly dispersion-free characteristics. The flat-band voltages of the three heterostructures are extracted to be 1.21 V, 1.50 V, and 1.29 V, respectively. The band diagrams of the three heterostructures were constructed based on the measured flat-band voltages, as shown in Figures 5d, 5h, and 5l, respectively. The C-V, reverse bias current, and breakdown characteristics of the three additional types of diodes beyond the Ge/Si diode have further proven the legitimacy of the ultra-thin oxide interface approach for creating lattice-mismatch heterostructures.

**Conclusions**

In summary, we demonstrated the highly-desired lattice-mismatched heterostructures by employing ultrathin oxide as an interface layer. It is anticipated that, based on the technical approach, a much larger capacity for material combinations than the lattice-matched ones can be further explored for prospective fabrication of a myriad of newfangled devices and systems. In the near future, solving fundamental materials challenges (*e.g.* doping) in wide bandgap semiconductors using this approach may be worth exploring.

**Acknowledgements**

The work was supported by the following grants in their chronological order: AFOSF PECASE (FA9550-09-1-0482, PM: Dr. Gernot Pomrenke), ONR (N00014-12-1-0077, PM: Dr. Daniel

**Author Contributions**

All performed the research. Z.M. conceived the idea, and designed and managed the entire research. D.L. and Z.M. developed the theoretical model and analyzed the data. J.-H.S. and Z.M. developed the generic heterostructure fabrication methods. S.J.C., J.-H.S., K.K., and Z.M. fabricated and characterized the heterostructures and devices. D.L. and Z.M. wrote the paper. M. Kim and Z.M. fabricated the germanium-on-insulators. J.S., X.Y., and X.W. performed the TEM studies and the crystal structural analyses. M.A.B. and P.K.M. grew the n-type GaAs epilayer. W.C., C.Z, and X.L. grew the p-type GaAs epilayer. D.L. and J.-H.S contributed to the research design. D.L., J.K., J.P., S.M., J.H.R., and S.G. contributed to the fabrication and the characterizations of heterostructures and devices. Z.M., D.L., J.-H.S., and S.J.C. contributed to the TEM studies. I.K.L. contributed to the device characterizations. S.J.C and J.-H.S. contributed to the data analysis. J.G. contributed to the model development and data analyses. J.B. contributed to the TEM sample preparation. S.J.C and J.-H.S. contributed to the paper writing.

**Competing interests**

The authors declare no competing financial interests.

**Data availability**

The datasets generated and analyzed in this article are available from the corresponding author on reasonable request.

**Figure Captions:**

**Figure 1 | Illustration of physical principles of an ultrathin oxide-interfaced heterostructure p-n junction. a**, A heterostructure p-n junction with depletion regions is formed under the equilibrium state. In between the p- and n-type semiconductors, an ultrathin dielectric/interface layer is present. $W_p$ and $W_n$ are the depiction region widths of the p- and n-type semiconductors,

respectively. **b,** Charge distribution ($\rho(x)$) of the p-n junction. The charges consist of ionized donors, $Q_{scn}$, ($Q_{scn} = qN_dW_n$) and acceptors, $Q_{scp}$, ($Q_{scp} = qN_aW_p$) in the respective depletion regions, and the surface charges located on the respective sides of the junction, $Q_{ssn}$ and $Q_{ssp}$. It is arbitrarily assumed that a net positive surface charge ($Q_{ssp}$) is on the p type semiconductor surface and a negative surface sheet charge ($Q_{ssn}$) is on the n type surface. **c,** Electric field ($E(x)$) distribution of the p-n junction. **d,** Electrical potential ($\phi(x)$) of the p-n junction. $V_i$ (can be negative) is the voltage drop across the interface layer. Note: the black dashed narrow box in **b**, **c** and **d** only illustrates the position of the interface layer. The generic illustrations are used in **b**, **c** and **d**. The following illustrations using p+/n- Ge/Si as the example. **e,** The band alignment of a type-II heterojunction of two semiconductors before contacting each other. Note: the bandgap of the interface layer is not scaled with respect to that of the semiconductors. **f,** Band diagram under the equilibrium state with the two semiconductors in contact via the interfacial layer in between. $q\psi_p$, $q\psi_n$ and $qV_i$ are the built-in electrical potentials carried by the two semiconductors and the interface layer, respectively. **g,** Band diagram of the p-n junction with a forward bias voltage $V$ applied. Under the applied voltage $V$, the electrical potentials carried by the two semiconductors and by the interface layer are $q\psi'_p$ ($= q\psi_p - \Delta q\psi_p$), $q\psi'_n$ ($= q\psi_n - \Delta q\psi_n$), and $qV'_i$ ($= qV_i - \Delta qV_i$), respectively. The potential changes from their equilibrium state comprise the applied voltage $V$ ($= \Delta\psi_p + \Delta\psi_n + \Delta V_i$). **h,** A calculated graphic representation of normalized diode current density ($J/J_s$, $J_s$ is saturation current) in its logarithmic scale, as a function of normalized voltage ($qV/kT$) with varied interface states density ($D_{ss}$). As $D_{ss}$ decreases, the diode current trend gets closer to an ideal diode (the pink line, $n = 1$), exhibited by the decreasing ideality factor, until the resistance-limited current region is reached (a series resistance of $10^{-14}/J_s$ $\Omega\cdot cm^2$ was artificially added). It is noted that the Shockley-Read-Hall (SRH) process-induced recombination current (i.e., $n = 2$) is not considered in the calculation.

**Figure 2 | Ge/Si heterostructure with ultrathin Al$_2$O$_3$ as the interface layer. a,** Schematic of a p+/n- Ge/Si heterostructure formed with ultrathin Al$_2$O$_3$ as its interface along with thickness and doping concentration values labeled for each layer. A 250 nm thick single crystalline Ge membrane was chemically bonded to Al$_2$O$_3$-covered single crystal Si surface. The detailed fabrication process is shown in Fig. S3. **b,** A false-colored SEM image of a finished diode fabricated from the Ge/Si heterostructure. The diameter of the Ge mesa is 10 μm. The left inset shows an optical image of

the diode after finishing anode and cathode metals but before the interconnect metal was formed. The anode metal was formed directly on the p+ Ge layer. The cathode metal was formed on the heavily doped n+ Si layer after mesa etching. **c,** An HRTEM image of the interface region of the Ge/Si heterostructure. The interface Al$_2$O$_3$ in this specifically examined heterostructure via HRTEM was deposited for 5 cycles. The single crystallinity of both Ge and Si were maintained. The insets are the diffraction patterns of the individual bulk materials with their zone indexing. The Al$_2$O$_3$ layer is clearly shown in the HRTEM image by the light-colored region. It is worth noticing that the Al$_2$O$_3$ layer shows discernable periodic patterns, which indicates strongly bonded interfaces and high-quality surface passivation. **d,** Atomic-scale illustration of the Ge/Al$_2$O$_3$/Si interface region to emulate the information shown in **c**. The passivation effect is mainly attributed to the formation of strong Si-O and Ge-O bonding, despite of the large lattice mismatch along the interfaces. Note that the Al$_2$O$_3$ atomic arrangement is not aligned with either the Si or the Ge lattices.

**Figure 3 | Electrical characteristics of Ge/Si heterostructure diodes.** Measured current density-voltage (*V*) results of p+/n- Ge/Si heterojunction diode **a,** without Al$_2$O$_3$ layer, **b,** with 3 cycles, **c,** 5 cycles, **d,** 7 cycles, and **e,** 10 cycles of Al$_2$O$_3$ layer. The multiple curves shown in **a**-**e** were obtained from multiple diodes selected in the representative areas of each sample. **f,** The calculated ideality factor values (*n*) and corresponding interface states density values ($D_{ss}$) for each tested Al$_2$O$_3$ thickness along with the means and standard deviations. **g,** C-V measurements of the diode with 5 cycles ALD Al$_2$O$_3$ under frequencies of 10 kHz, 100 kHz, and 1 MHz. The extracted built-in potential is 0.55 V. **h,** Constructed band diagram under equilibrium based on the extracted built-in potential showing a voltage drop of -0.09 V across the interface layer, $\Delta \psi_p + \Delta \psi_n - 0.09$ V = 0.46 V (see Fig. S8). **i,** Measured temperature-dependent reverse bias and breakdown characteristics. The breakdown voltage increased from -26.5 V, to -27 V and -29 V as the temperature was elevated from 25 °C, to 50 °C, and 75 °C, respectively. Abrupt breakdown characteristics along with a positive temperature dependence of the breakdown voltages are the indications of occurrence of avalanche breakdown.

**Figure 4 | Si/GaAs, GaAs/GaN and Si/GaN heterostructures and their p-n diodes.** Note: The interface $Al_2O_3$ in all the three heterostructures was deposited for 5 cycles. An HRTEM image of the interface region of the **a,** Si/GaAs, **e,** GaAs/GaN, and **i,** Si/GaN heterostructures. The $Al_2O_3$ layer in each of the heterostructures is shown by the light-colored regions in the three HRTEM images. The insets in **a, e**, and **i**, are the diffraction patterns of the individual bulk materials that form the heterostructures with their zone indexing. Measured current density versus voltage characteristics of **b**, Si/GaAs **f**, GaAs/GaN, and **j**, Si/GaN p+/n- diodes. The ideality factors of the three diodes are 1.07, 1.13, and 1.14, respectively. The insets in **b**, **f**, and **i**, show the respective diode schematics with their layer thicknesses and layer doping concentrations labeled. C-V characteristics of **c**, Si/GaAs, **g**, GaAs/GaN, and **k**, Si/GaN p+/n- diodes measured at 1 MHz. The extracted built-in potential from the three C-V curves are 1.21 V, 1.50 V, and 1.29 V, respectively. Constructed band diagram of **d**, Si/GaAs **h**, GaAs/GaN, and **l**, Si/GaN p+/n- junctions under the equilibrium state, $q\psi \approx q\psi_n$. The voltage drop values across the interface layer for the three diodes are -0.14 V, -0.13 V, and -0.23 V, respectively.

**Extend Data Figure 1 | Electrical characteristics of Si/GaAs, GaAs/GaN, and Si/GaN heterostructure diodes under reverse bias.** Measured current density (Left Y- axis, linear scale. Right Y-axis, logarithmic scale) as a function of voltage under reverse bias at room temperature for **a,** p+/n- Si/GaAs, **b,** p+/n- GaAs/GaN, and **c,** p+/n- Si/GaN diodes. The breakdown voltages of the three diodes are -17.7 V, -30.6 V, and -28.3 V, respectively.

**Extend Data Figure 2 | C-V characteristics of Si/GaAs, GaAs/GaN, and Si/GaN heterostructure diodes under different frequencies.** Measured C-V characteristics of **a**, Si/GaAs, **b**, GaAs/GaN, and **c**, Si/GaN p+/n- diodes at 10 kHz, 100 kHz, and 1 MHz.

**Extend Data Table 1 | Comparison of key diode performance parameters, ideality factor (*n*) and $I_{ON}/I_{OFF}$ ratio, between oxide-interfaced heterostructure diodes and diodes fabricated employing other methods.**

**METHODS**

Here we describe in detail the growth of epitaxial substrates, the preparation of single crystalline Ge, Si, and GaAs membranes, and fabrication of Ge/Si, Si/GaAs, GaAs/GaN, and Si/GaN heterostructures and their p+/n- diodes. We also describe the characterizations of the heteterostructure diodes.

**Growth of epitaxial substrates.** The n-/n+ (500 nm/400 nm) Si epilayer (Fig. 2a) was purchased from Lawrence Semconductor Research Lab. The n-/n+ (500 nm/500 nm) GaAs epilayer (the inset of Fig. 4b) was grown using a metalorganic chemical vapor deposition (MOCVD) system. The n-/n+ (130 nm/1.45 μm, the insets of Fig. 4f and 4j) GaN layer was purchased from Cree, Inc.

**Fabrication of Ge/Si heterostructure and p+/n- diodes**. Germanium-on-insulator (GeOI) wafers of 4" in diameter were fabricated using previously described procedures (Ref. 56). The boron doping concentration ($2\times10^{18}$ cm$^{-3}$) of the 250 nm thick p+ Ge layer inherited from source Ge wafers used for GeOI wafer fabrication. The Ge membrane release process started with patterning the top Ge layer with etching holes of $5\times5$ μm$^2$ via a dry etching process using a reactive PT790 RIE plasma etcher with gas flow rates of SF$_6$: 67 sccm and O$_2$: 5 sccm, pressure of 15 mTorr, and power of 100 W. The top Ge layer was then released by etching away the buried oxide layer (BOX: 150 nm thick) in a diluted hydrogen fluoride (HF: H$_2$O) solution (1:1) for 1.5 hours. The released Ge membrane sitting on the Si handling substrate was ready for transfer after drying.

The n-/n+ Si epilayer (Fig. 2a) substrate was cleaned following standard cleaning procedures. Native oxide on the wafer surface was removed using diluted (1:10) HF solution followed by a thorough deionized water rinse. The sample was dried in a nitrogen environment and loaded into a nitrogen-filled glove box integrated with an Ultratech/Cambridge Nanotech Savannah S100 atomic layer deposition (ALD) system. The ALD chamber was pre-heated to 200°C and pumped down to vacuum (<0.1 mTorr) after the sample was loaded. During the ALD process, trimethylaluminum (TMA) and water vapor were pulsed separately into the ALD chamber for 0.015 second each, and separated by 5 second nitrogen gas purging. The Al$_2$O$_3$ growth rate was ~0.1 nm per cycle. After ALD deposition, the released p+ Ge membrane was transferred to the Al$_2$O$_3$-coated Si epilayer using a polydimethylsiloxane (PDMS) stamp (Ref. 57). It is noted that the PDMS-based transfer techniques are capable of high throughput for industrial-scale manufacturing (Ref. 58). Then a rapid thermal anneal (RTA) procedure at 250°C for 5 min followed to form chemical bonding. Raman spectroscopy was performed on the sample.

The p+/n- diode fabrication using the heterostructure began with anode patterning on top of the p+ Ge layer using photolithography, electron-beam evaporation of Ni/Au (10/100 nm), and a lift-off process. Using the anode metal as a mask, the p+ Ge and n- Si layers were etched using the PT790 RIE plasma etcher with gas flow rates of $SF_6$: 8 sccm and $O_2$: 1 sccm, pressure of 6 mTorr, and power of 30 W until the n+ Si layer was exposed. Cathode metals of Ti/Au (10/100 nm) were formed on the n+ Si layer using the same method as the anode metal. Covering the diode area with photoresist, the n+ Si layer was etched away using RIE to reach the Si substrate for device isolation. A 500 nm $SiO_2$ layer was deposited using plasma-enhanced chemical vapor deposition (PECVD) for device passivation, followed by via opening using photolithography and RIE ($CF_4$: 45 sccm, $O_2$: 5 sccm, pressure: 40 mTorr, and power: 100 W), interconnection metal deposition (Ti/Al/Ti/Au: 15/650/15/100 nm), and lift-off.

**Fabrication of Si/GaAs heterostructure and p+/n- diodes.** Silicon-on-insulator (SOI) wafers with 205 nm thick p-type (boron, 11.5 ohm-cm) top Si layer were acquired from Soitec and first converted to p+ SOI. A screen oxide of 30 nm thick was thermally grown on SOI wafers with 25 nm Si consumed. A boron ion implantation was performed by CuttingEdge Ions, LLC using a 7-degree angle of incidence at a dose of 4 x$10^{15}$ $cm^2$ and energy of 15 keV, followed by thermal anneal at 950 °C for 40 min in nitrogen ambient for Si recrystallization and boron dopant re-diffusion. Details of the procedures can be found elsewhere (Ref. 59). The resulting doping concentration across the 180 nm thick top Si layer is $5\times10^{19}$ $cm^{-3}$. Similar to the Ge membrane release, Si membrane was ready after forming etching holes, etching away the BOX layer in dilute HF (1:1) for 1 hour, and drying.

The n-/n+ GaAs epi layer was cleaned following standard cleaning procedures for GaAs. Native oxide was removed by immersing the sample in $NH_4OH:H_2O$ (1:5) and diluted (1:10) HF solutions each for 1 minute. After the sample was loaded into an ALD chamber, TMA precursor flow for five cycles was used to remove native oxide, followed by five cycles of $Al_2O_3$ deposition. The released p+ Si membrane was transferred to the $Al_2O_3$-coated n-/n+ GaAs using the same method as that used for Ge membrane. An RTA procedure was carried out at 350°C for 5 min in a nitrogen environment to finish the heterostructure fabrication. Using the similar method as Ge/Si diode fabrication, p+/n- Si/GaAs diodes were fabricated. The anode metals (Ti/Au: 20/200 nm) were formed on the p+ Si layer. After etching away the p+ Si layer using the PT790 RIE plasma etcher ($SF_6$: 67 sccm, $O_2$: 5 sccm, pressure: 15 mTorr, and power: 100 W) and the n- GaAs layer using an inductive coupled plasma (ICP) etcher (PT770ICP, $BCl_3$/Ar: 10/5 sccm, pressure: 20 mTorr, plasma power: 56 W, and inductor power: 500 W) with the anode metal as a mask, the n+ GaAs layer was exposed. Cathode metals (Pd/Ge/Au: 30/40/200 nm) were formed on the n+ GaAs layer, followed by a thermal anneal at 400℃ for 30 seconds in an ambient forming gas ($H_2$/$N_2$: 5%/95%) to form an ohmic contact. The diodes were passivated by coating a 8 nm thick $Al_2O_3$ using ALD at 250 ℃.

**Fabrication of GaAs/GaN heterostructure and p+/n- diodes.** The 100 nm epitaxial p+ GaAs layer (Fig. 4f) was grown on top of a sacrificial $Al_{0.95}Ga_{0.05}As$ layer which was in turn grown on a GaAs substrate using MOCVD. After forming etching holes using photolithography and ICP etching, the p+ GaAs membrane was released by immersing the sample in a diluted HF (1:100) solution for 10 minutes to etch away the $Al_{0.95}Ga_{0.05}As$ layer. The residue, $AsH_3$ and either $AlF_3$ or $Al(H_2O)_n^{3+}$, [Ref. 60] formed on the back side of the released GaAs membrane was removed using tetramethylammonium hydroxide (TMAH)-based solution (Microchem MIF-321) after the membrane was picked up with a PDMS stamp. Crack-free GaAs membranes with clean surfaces were fabricated and ready for transfer.

To fabricate p+/n- GaAs/GaN diodes, the n- GaN layer of the n-/n+ GaN epilayer substrate was selectively etched with photoresist as a mask to protect the remaining part of the n- GaN, using an ICP etcher ($BCl_3$/$Cl_2$/Ar: 10/16/3 sccm, pressure: 4 mTorr, plasma power: 100 W, and inductor power: 500 W) until the desired regions of the n+ GaN layer were exposed. After dry etching, the photoresist was stripped off. Following standard cleaning procedures and a native oxide-removal procedure, the samples were loaded into the ALD chamber where $Al_2O_3$ was deposited for five cycles. Cathode metals (Ti/Al/Ti/Au: 20/100/45/250 nm) were then formed on the exposed n+ GaN layer, followed by thermal annealing at 600℃ for 30 seconds and 800℃ for 30 seconds in ambient forming gas ($H_2$/$N_2$: 5%/95%). The PDMS-picked GaAs membrane was then transferred to the top of the n- GaN regions (previously protected with photoresist), followed by thermal annealing at 400 ℃ for 5 minutes to form chemical bonding. Anode metals (Pt/Ti/Pt/Au: 10/40/10/250 nm) were then deposited on the p+ GaAs layer. A thermal anneal at 375℃ for 40 seconds was carried out in ambient forming gas ($H_2$/$N_2$: 5%/95%) to form ohmic contact.

**Fabrication of Si/GaN heterostructure and p+/n- diodes.** The Si/GaN heterostructure p+/n- diode fabrication process is identical to that of the GaAs/GaN diodes. The Si membrane used here was the same as the one used in Si/GaAs heterostructure fabrication. The thermal annealing procedure to form chemical bonding between p+ Si and n- GaN was carried out at 500℃ for 5 minutes. The anode metals (Ni/Au: 10/200 nm) were deposited on Si.

**Electrical characterizations of the diodes.** The I-V characteristics of the diodes were measured using a Keithley 4200-SCS semiconductor characterization system at room temperature (RT) and elevated temperatures by heating the sample stage. C-V characteristics of the p+/n- junctions were measured using Keysight Technologies E4980A precision LCR meters at RT.

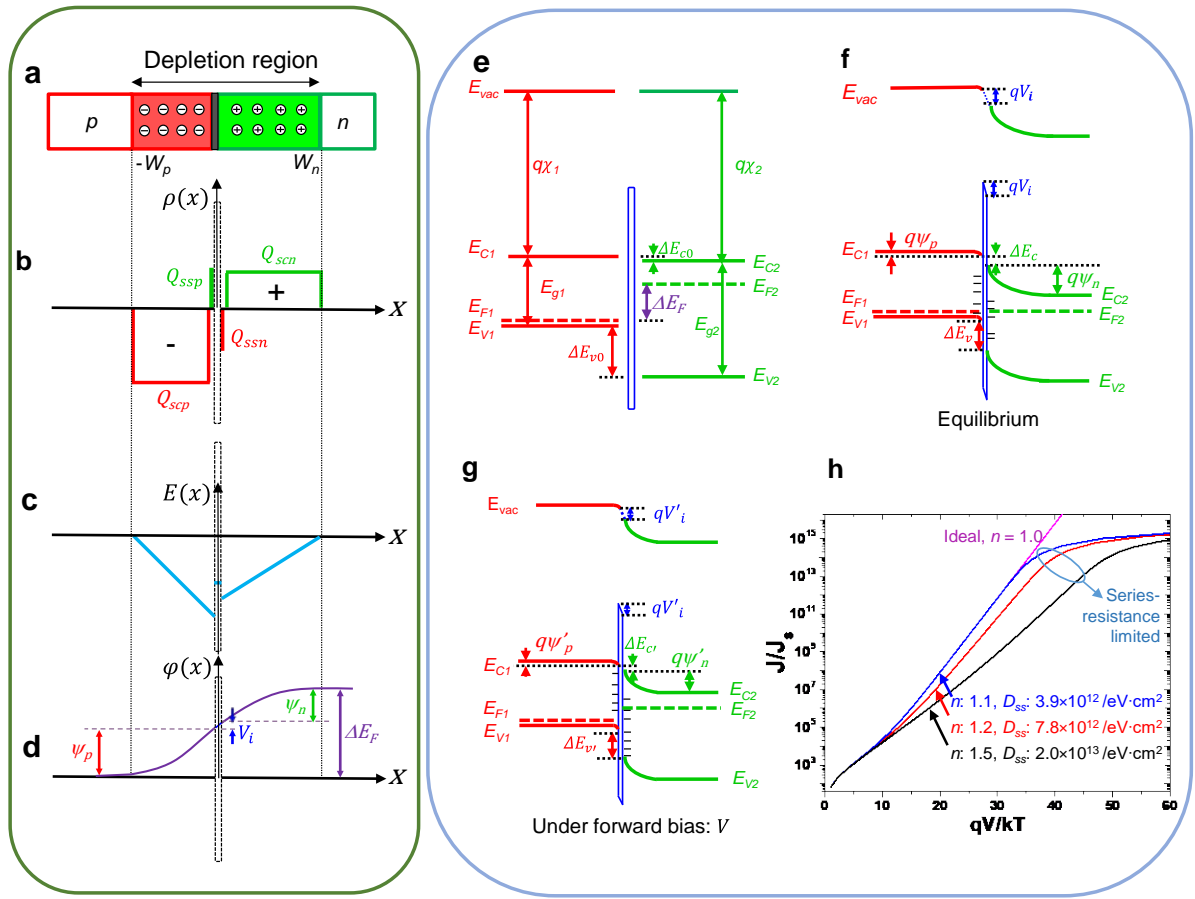

**Figure 1**

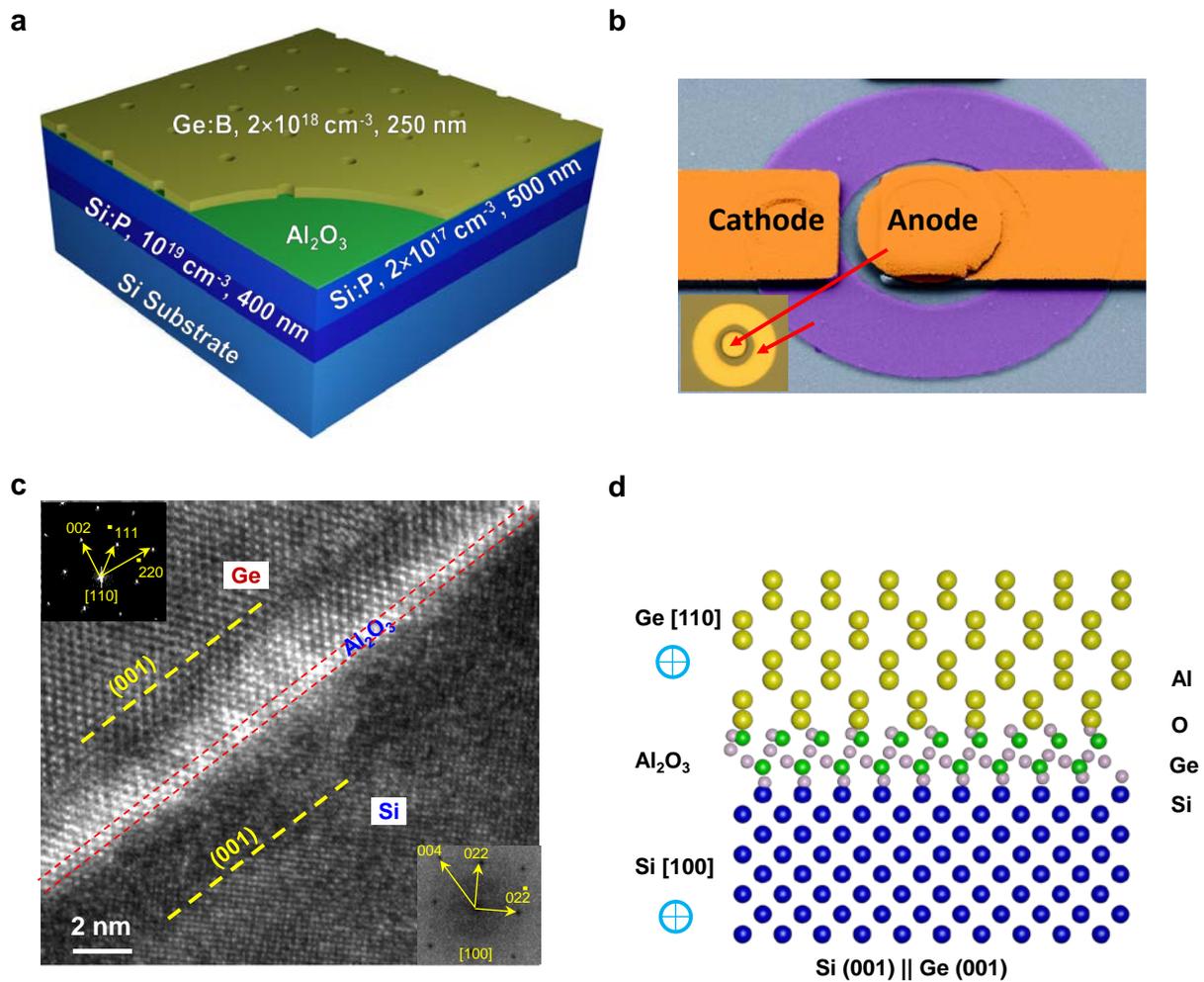

**Figure 2**

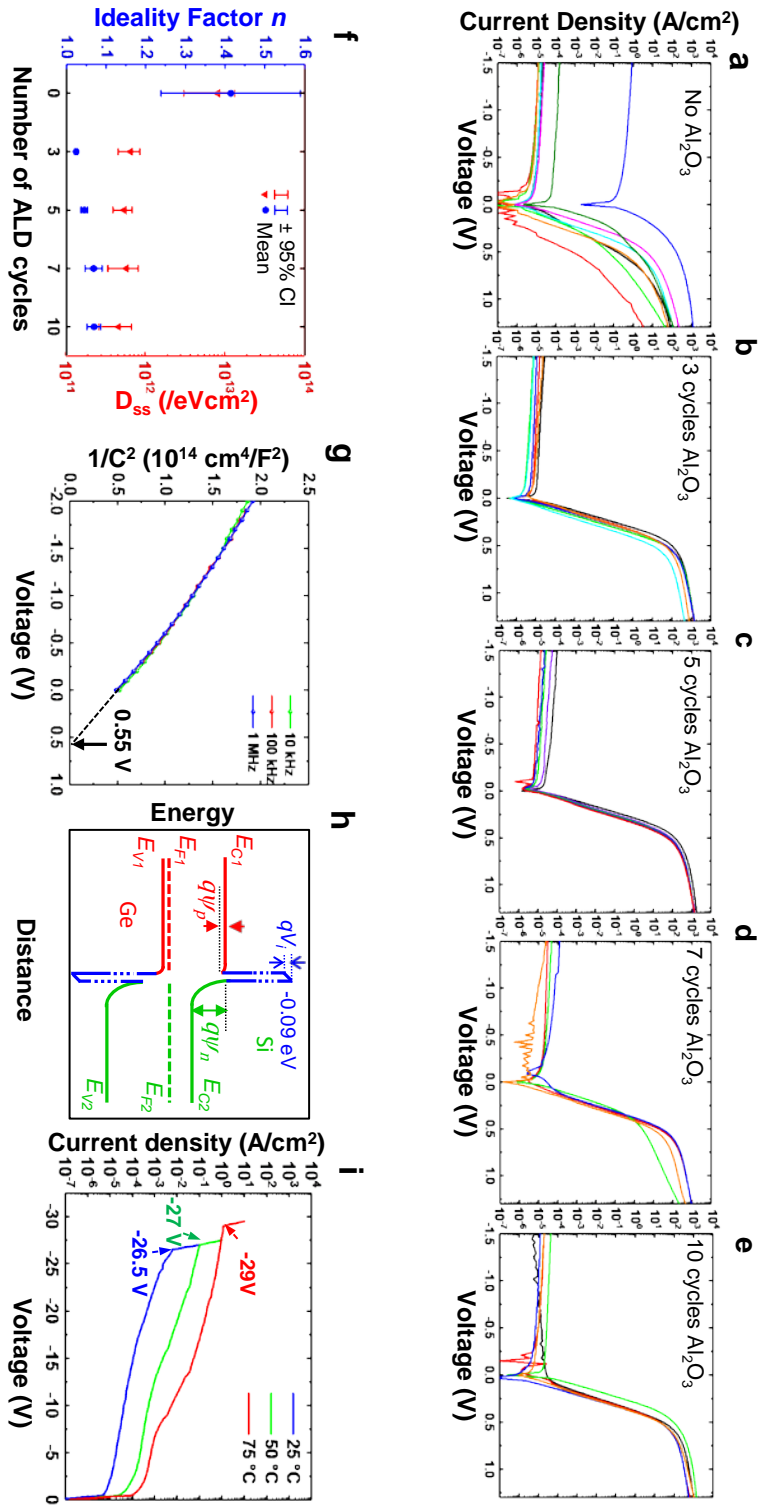

**Figure 3**

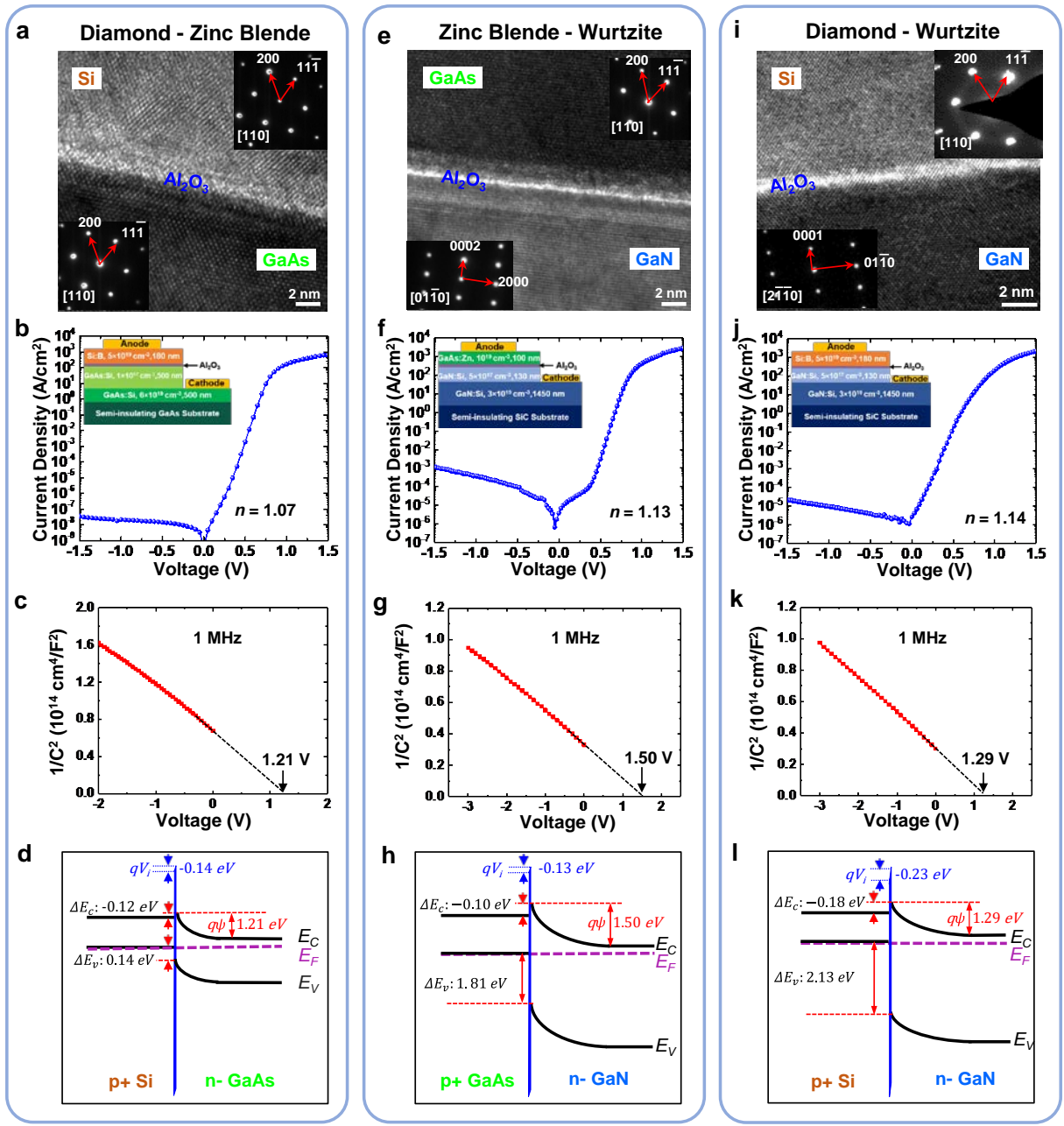

**Figure 4**

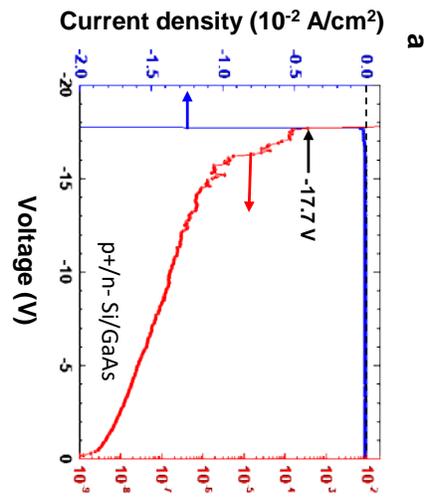
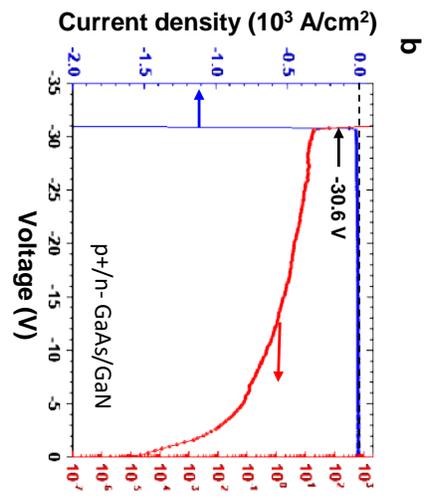
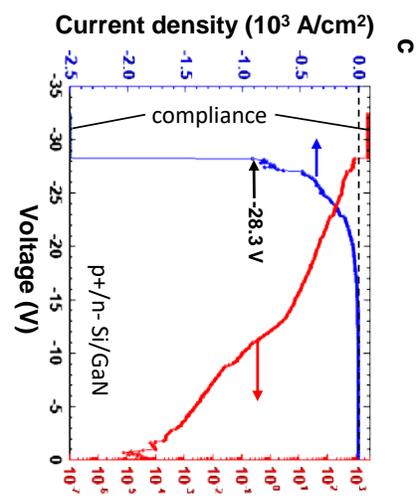

**Extended Data Figure 1**

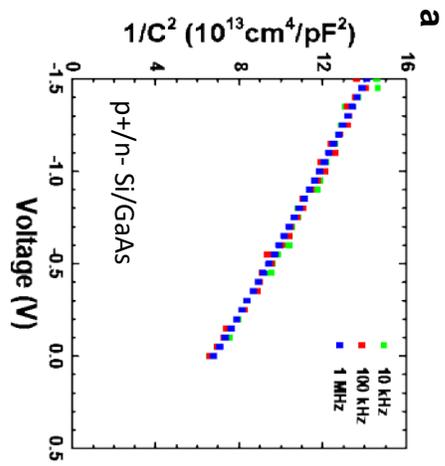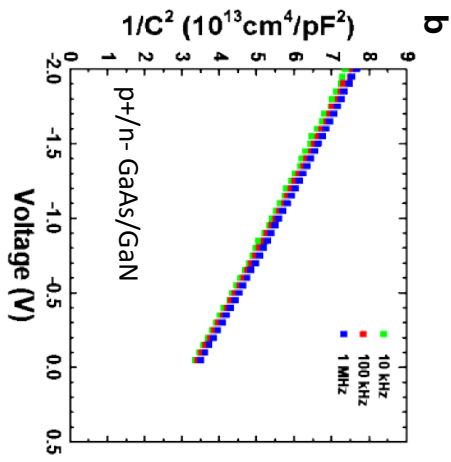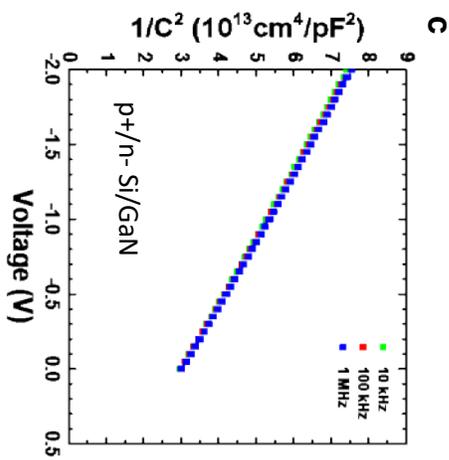

**Extended Data Figure 2**

| Heterojunction | Method | Ideality Factor ($n$) | $I_{ON}/I_{OFF}$ ratio (@ $\pm$1V) | Reference |
|---|---|---|---|---|
| p-Ge/n-Si | Selective heteroepitaxy | | ~$10^5$ | [37] |
| p-Ge/n-Si | Selective heteroepitaxy with multiple hydrogen anneals | - | ~$2.43\times10^3$ | [38] |
| p-Ge/n-Si | Heteroepitaxy of Ge on Si | 1.1 | ~$5\times10^7$ | [39] |
| n-Ge/p-Si | Heteroepitaxy of Ge on Si | 1.16 | ~$3\times10^6$ | [39] |
| p-Ge/n-Si | Low temperature wafer bonding | 2.28 | ~$8.5\times10^2$ | [40] |
| p-Ge/n-Si | Micro bonding a Ge beam on Si | - | ~$7.5\times10^3$ | [41] |
| p-Ge/n-Si | Low temperature wafer bonding | - | ~$7.5\times10^1$ | [42] |
| p-Ge/n-Si | Ribbon bonding | 2.15 | ~$2.0\times10^2$ | [43] |
| p-Ge/n-Si | **Direct bonding (no $Al_2O_3$)** | **1.41$\pm$0.23** | **~$10^{5\pm1}$** | **This work** |
| p-Ge/n-Si | **$Al_2O_3$-interfaced bonding** | **1.02** | **$1.5\times10^8$** | **This work** |
| p-Si/p-GaAs | Wafer bonding | 1.5 | ~$1\times10^3$ | [50] |
| n-Si/p-GaAs | Surface activated bonding | - | ~$2.0\times10^2$ | [53] |
| p-Si/n-GaAs | Surface activated bonding | - | ~$2.7\times10^3$ | [53] |
| n-Si/p-GaAs | Molecular beam epitaxy | 1.5 | ~$1\times10^6$ | [54] |
| p-Si/n-GaAs | **$Al_2O_3$-interfaced bonding** | **1.07** | **$7.9\times10^9$** | **This work** |
| p-GaAs/n-GaN | Wafer bonding | ~1.83 | ~$4\times10^3$ | [51] |
| p-GaAs/n-GaN | Wafer fusion | 1.75 | ~$1\times10^4$ | [52] |
| p-GaAs/n-GaN | **$Al_2O_3$-interfaced bonding** | **1.13** | **$1.1\times10^6$** | **This work** |
| p-Si/n-GaN | Molecular beam epitaxy | ~1.5 | ~$1\times10^3$ | [55] |
| p-Si/n-GaN | **$Al_2O_3$-interfaced bonding** | **1.14** | **$4.1\times10^8$** | **This work** |

**Extended Data Table 1**